# Rain Attenuation Prediction for Terrestrial Microwave Link in Bangladesh

HOSSAIN Sakir

International Islamic University Chittagong, Bangladesh,
Department of Electronic and Telecommunication Engineering, Faculty of Science and Engineering,
4314 Kumira, Chittagong, Bangladesh, Email: shakir.rajbd@yahoo.com

*Abstract* - **Rain attenuation is a major shortcoming of microwave transmission. As a subtropical country, Bangladesh is one of the highest rainy areas of the world. Thus, designing a terrestrial microwave link is a serious challenge to the engineers. In this paper, the annual rain rate and monthly variation of rate are predicted for different percentage of time of the year from the measured rainfall data. Using ITU rain model for terrestrial microwave communication, the rain attenuation is predicted for five major cities of Bangladesh, namely Dhaka, Chittagong, Rajshahi, Sylhet, and Khulna. It is found that rain attenuation is the most severe in Sylhet and least in Rajshahi. The attenuation is estimated for different frequency and polarization. A horizontally polarized signal encounters 15% more rain attenuation than that of vertically polarized signal. It is also found that attenuation in Rajshahi is about 20% lesser than that in Sylhet. Thus, the horizontally polarized transmission in Rajshahi experiences about 5% less attenuation than the vertically polarized transmission in Sylhet.**

## I. INTRODUCTION

The usage of wireless communication is following a sharp uptrend. People are passing the busiest time of human civilization. So, installing optical fiber or coaxial cable is being seen very time consuming and costly. Even personal or cooperate communication between two near buildings is not done by cable. Mobility is an important precondition of today's communication systems. This is why; besides long haul communication, terrestrial microwave communication is being used for short distance communication. If there are five offices of a company in a city, the use of cable connection is going down considerably to connect LANs of these offices. Instead, the short-distance point-to-point connection is established between different LANs using microwave. Even for connecting remote CCTV to the LAN, microwave link is being used. Thus, the terrestrial microwave communication will be used extensively in future. However, the problem that engineers face to design a microwave link is the rain attenuation, because microwave signal is very much affected by the rain drops. When a microwave signal comes in contact with the rain drop, the signal strength gets reduced due to the absorption and scattering of energy of the signal by the rain drop. Moreover, the rain drop changes the polarization of the signal. These entire phenomenons cause the attenuation of microwave signal, and this attenuation rises rapidly with the slight increase of frequency above 7 GHz and the precipitation rate. But, to meet the growing demand of microwave frequency for terrestrial communication, studying the feasibility of the use of upper band frequency is a pre-requisite task. One of the most important tasks is to investigate the effect of rain on the performance of terrestrial microwave link and thereby estimating the amount of link margin needed to compensate this adverse effect. This rain attenuation is largely dependent on the amount of precipitation occurs in a particular region. Such estimation has already been done for a number of countries like India, Malaysia, South Africa and so on [1]-[3]. Bangladesh, as a sub-tropical country, is one of the rainiest regions of the world. So, the feasibility of using upper band of microwave frequency is an important matter of investigation. In this paper, the rain rate is computed for different regions of Bangladesh based on the measured data, and the corresponding rain attenuation of terrestrial microwave signal is estimated and analyzed from different perspectives.

## II. RAIN RATE DISTRIBUTION IN BANGLADESH

Bangladesh has a subtropical monsoon climate characterized by wide seasonal variations in rainfall. Heavy rainfall is a feature of Bangladesh. There is significant variation in annual rainfall across Bangladesh; this is highest at Sylhet due to its location in the south of the foothills of the Himalayas, while the minimum is experienced at Rajshahi, relatively dry northern region of Bangladesh. Including the rain attenuation in the mentioned two cities, the same for Dhaka, Chittagong, and Khulna are predicted here. The outage of a terrestrial microwave system depends considerably on the signal attenuation due to rain, which in turn depends on instantaneous rain rate. But the rain rate is not constant over the year. So the Cumulative Distribution (CD) of rainfall across the year is used to predict rain attenuation. Here the annual rainfall statistics of forty years (from 1968 to 2008) of five different cities are collected from the website of Bangladesh Agriculture Research Council (BARC) [4] and the average annual rainfall is computed from that data. To get the CD from the annual rainfall amount, there are number of rain rate prediction models including Rice-Holmberg Rain model (R-H model) [5], Moupfouma Rain model [6]. Of them, R-H model can provide a good estimation of one hour rain rate for 1-minute integration of time from the annual precipitation rate in the Bangladeshi climatic condition [7]. This is why; R-H model was used here to estimate that. The parameters needed for R-H model include Thunderstorm Ratio β (amount of convective rain compared to the total rainfall) and the average annual rainfall depth M. The values of these parameters are shown in Table I, where β is taken from the map in [5].

TABLE I: Parameters of R-H rain model

| Parameters | Dhaka | Chittagong | Rajshahi | Sylhet | Khulna |
|---|---|---|---|---|---|
| M (mm) | 2124 | 2887 | 1545 | 4101 | 1819 |
| β | 0.5 | 0.5 | 0.5 | 0.5 | 0.5 |

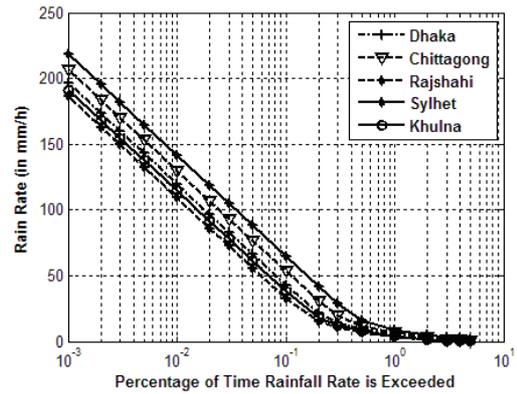

Fig. 1. Cumulative rain distribution of Bangladesh by R-H rain model

The R-H model gives the rain rate CD at 1-minute integration time which is shown in Fig. 1 for the parameters mentioned in Table I. The 0.01% rain rates ($R_{0.01}$) of four cities taken from Fig. 1 are listed in Table II, where the maximum $R_{0.01}$, which is 141.661 mm/h, is reported in Sylhet, while the minimum of 109.1496 mm/h in Rajshahi. However, the predicted 0.01% rain rate in Bangladesh is 95 mm/h according to Recommendation ITU-R P.837-6 [8]. Since the recommendation in [8] predicts the rain rate of all regions of the world, it can not accurately predicts 0.01% rain rate for Bangladesh. A significant difference of 12% is observed between the measured $R_{0.01}$ and ITU predicted $R_{0.01}$. The month-to-month variation of the rainfall of different cities of Bangladesh are presented in Fig. 2, the maximum precipitation is reported in July and the minimum in January.

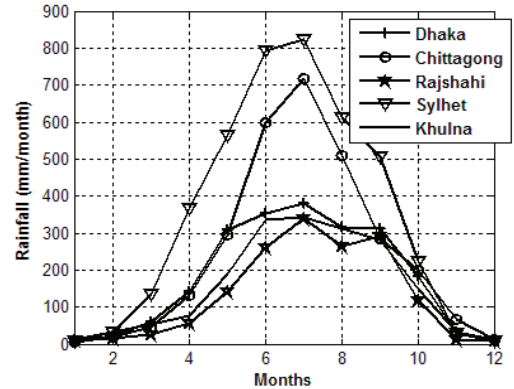

Fig. 2. Rainfall amount across the year.

TABLE II: $R_{0.01}$ for different cities

| City | $R_{0.01}$(mm/h) |
|---|---|
| Dhaka | 119.7673 |
| Chittagong | 129.9933 |
| Rajshahi | 109.1496 |
| Sylhet | 141.6991 |
| Khulna | 114.6028 |
| ITU Map | 95 |

III. RAIN ATTENUATION PREDICTION MODEL

The rain attenuation of terrestrial microwave link depends on the amount of rain, along with the size of the rain drops, which intersects the link. But the rain drops' amounts and sizes vary temporally and spatially. This is why, measuring these two parameters in a part of the path is not sufficient to predict the attenuation for the whole link path. Instead, Some techniques are needed to cope with the temporal and spacial variation of rain characteristics.

A number of rain attenuation prediction models are developed to overcome this difficulty. Each of these models proposes different techniques to predict the rain attenuation of the link path. Few examples of rain attenuation prediction models for terrestrial microwave prediction are ITU model [9], Moupfouma model [10], Lin model [11], and Silva Mello model [12]. All of these models follow the same steps. Firstly, compute the 0.01% of rain rate (mm/h) for 1-minute integration of time using Rice-Holmberg model [5] or Moupfouma rain rate model [6], secondly, compute the specific attenuation (db/km), and finally figure out the rain attenuation computing the effective path length of rain ($d_{eff}$). The above mentioned models follow the first two steps similarly, but differ in the third steps in computing effective path length. However, according to [2], the prediction of ITU model closely follows the measured rain attenuation for tropical region, like Malaysia. So, among these four models, ITU model has been used in the investigation of rain attenuation of terrestrial microwave signal in different cities of Bangladesh, because Bangladesh, as a subtropical country, has considerable similarity with the annual precipitation of Malaysia. The steps of ITU model [9], which is valid at least for frequency up to 40 GHz and lengths up to 60 km, are given below:

*Steps 1*: Compute the rain rate $R_{0.01}$. Such value can be computed using [5]. However, if such information is not available locally, an estimated $R_{0.01}$ can be found from the global map of rain rate exceeded for 0.01% of time using [5].

***Step 2:*** Compute the attenuation per kilometer, called specific attenuation, $\gamma_R$ (db/km), from $\gamma_R = kR^\alpha$, where the values of the constant k and α are originally obtain from [13]. The expression for the calculation of these parameters can be found in [14] for different frequencies and polarizations.

***Step 3:*** Though the specific attenuation gives the attenuation that the microwave link encounters in each kilometer, this parameter is not sufficient for computing the attenuation of the whole path, because there is a considerable temporal and spatial variation of the rain rate across the link. So, just multiplying the specific attenuation by the actual link path cannot properly give whole path attenuation. Instead, two new parameters are introduced in this model, namely rain cell length $d_o$, the length over which the rain is considered as uniform, and effective path length $d_{eff}$, the average length of the intersection between cell and link.

Effective path length, $d_{eff} = \dfrac{d}{1+\dfrac{d}{d_o}}$ (1)

where d is the actual length of the terrestrial microwave link and for $R_{0.01} \leq 100$ mm/h

$d_o = 35e^{-0.015 R_{0.01}}$ (2)

And for $R_{0.01} > 100$ mm/h

$d_o = 35e^{-1.5}$ (3)

***Step 4:*** An estimate of the path attenuation exceeded for 0.01% of the time is given by

$A_{0.01} = \gamma_R d_{eff}$ (4)

***Step 5:*** The rain attenuation exceeded for other percentage of time p in the range 0.001% to 1% can be computed from the following extrapolation formula:

For latitude $\geq 30°$

$A_p = 0.12 A_{0.01} p^{-(0.546 + 0.043 \log_{10} p)}$ (5)

And for latitude $<30°$

$A_p = 0.07 A_{0.01} p^{-(0.855 + 0.139 \log_{10} p)}$ (6)

***Step 6:*** The rain attenuation in the worst-month can be computed by calculating the worst-month time percentages $p_{wm}$ corresponding to the annual time percentages $p_a$ using the following formula [15]

$p_{wm} = e^{\dfrac{\ln(\dfrac{p_a}{0.3})}{1.15}}$ (7)

## IV. RAIN ATTENUATION PREDICTION

In this paper, the most important cities of Bangladesh were considered to find the rain attenuation, because almost 80% communication equipments are setup in these five cities, like Dhaka, Chittagong, Rajshahi, Sylhet, and Khulna. The rain attenuation of terrestrial microwave link depends on several factors, such as rain fall rates, polarization of the transmitted signal, frequency of the signal to be transmitted, and so on. The rain attenuation for terrestrial microwave communication will be investigated from these perspectives. The first task of this estimation process is to figure out the specific attenuation, that is, the attenuation in each kilometer in the rainy area. Fig. 3 and Fig. 4 show the specific attenuation of the mentioned five cities of Bangladesh for horizontal and vertical polarizations, respectively. The Fig. 3 shows that the signal attenuation increases sharply with frequency less than 50 GHz, followed by gradual rise between 50 to 100 GHz frequency, and remains almost constant for the rest of frequency. The reason behind this gradual increase in attenuation per km might be due to the loss of synchronization between the wavelength of signal and the rain drop size after 50 GHz. It is evident from Fig. 3 and Fig. 4 that the specific attenuation of vertical polarization is about 1dB lower compared to 40 dB specific attenuation of horizontal polarization at 100 GHz in Sylhet, which is the highest among the considered cities of Bangladesh, while the minimum of 33 dB is found at Rajshahi. In addition, the specific attenuation found from the measured data for Dhaka is about 10% higher than that found from the ITU predicted rain rate.

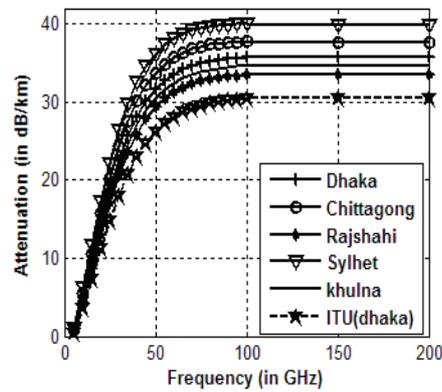

Fig. 3. Dependency of specific attenuation on frequency in horizontal polarization.

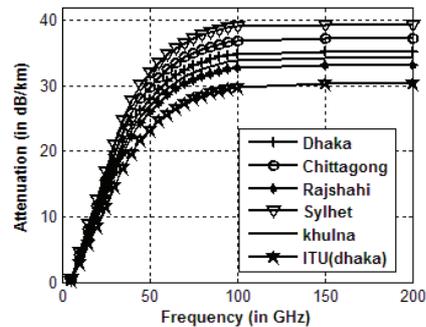

Fig. 4. Dependency of specific attenuation on frequency in vertical polarization.

Attenuation is related to the microwave link length. Fig. 5 shows the rain attenuation of five cities of Bangladesh for horizontal polarization for different distances. A sharp rise in attenuation is observed up to first 10 km, followed by gradual increase up to next 10 km, and then the attenuation becomes nearly flat for the next 40 km. The reason of this non-uniform attenuation change is that the effective path length does not remain so significant for large actual link length as it is for shorter link. The signal transmitted in horizontal polarization encounters rain attenuation of 27 dB in Sylhet for the first 10 km, while the minimum attenuation is observed in Rajshahi, where the figure is 20 dB, about 26% lower attenuation than that of Sylhet. Each city encounters about 24% lower attenuation in vertical polarization as shown in Fig. 6 compared to horizontal polarization.

Rain does not occur all times of year and its rate does not remain same all the time when it occurs. Thus, the amount of rain fade margin needed to compensate rain effect varies with time. Fig. 7 gives a graphical representation of rain attenuation that a horizontally polarized signal experiences for different percentage time of the year. For example, for $10^{-2}$ time of the year (i.e., 99.99% availability), the attenuation level at Sylhet is 40 dB, while it is 32 dB at Dhaka, and to ensure 99.9% availability, 15 dB and 12 dB fade margin are required for Sylhet and Dhaka. However, the amount of fade margin can be reduced up to 25 % by using vertical polarization as show in Fig. 8.

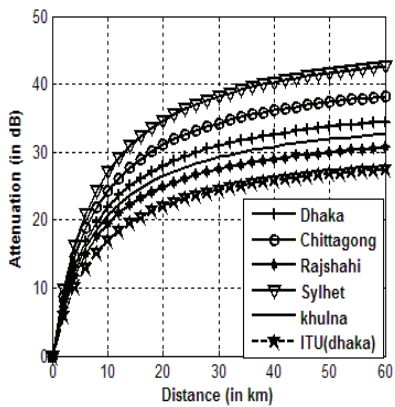

Fig. 5. Rain Attenuation for different distances at frequency of 10 GHz in horizontal polarization.

The maximum precipitation occurs in the month of June. So, determining attenuation is needed in the worst month to find the maximum rain fade margin required to continue the communication. Fig. 9 shows the worst month statistics for horizontal polarization. It is evident from Fig. 7 and Fig. 9 that the average annual fade margin required to ensure 99.999% availability is needed to ensure 99.99% availability in the worst month, that is, higher fade margin required in this month.

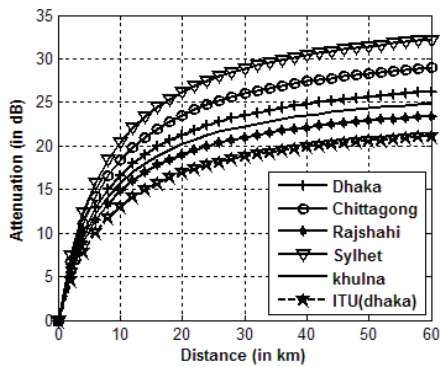

Figure 6. Rain Attenuation for different distances at frequency of 10 GHz in Vertical polarization.

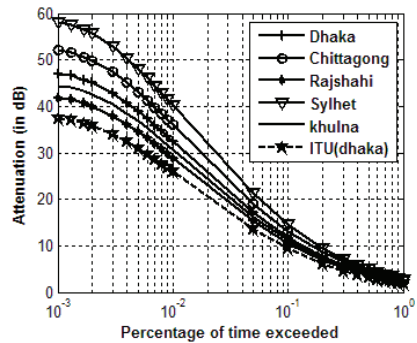

Fig. 7. Annual rain attenuation exceeded for different percentages of time at 10 GHz frequency with link length of 40 km in horizontal polarization.

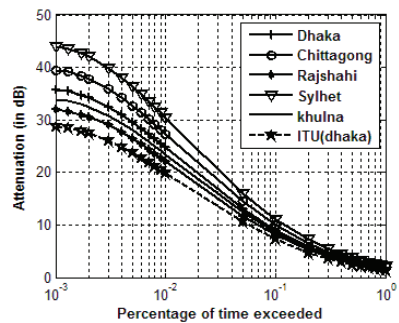

Fig. 8. Annual rain attenuation exceeded for different percentages of time at 10 GHz frequency with link length of 40 km using vertical polarization.

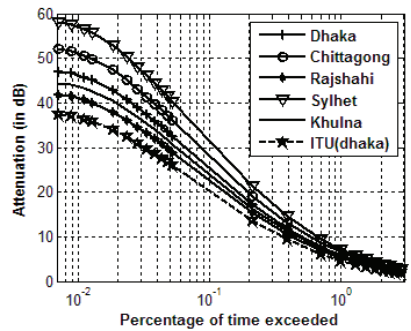

Fig. 9. Worst month rain attenuation at 10 GHz with link length of 40 km.

Rainfall does not remain same in all months; rather it varies heavily from month to month, as is revealed from Fig. 2. Thus, determining attenuation with respect to different months is needed to provide proper fade margin in different months of the year. Fig. 10 shows this attenuation variation in horizontal polarization at frequency of 10 GHz for ensuring 99.99% availability. The maximum rain attenuation of about 50 dB is observed for Sylhet and Chittagong and 40 dB in the rest of the cities. In contrast, the minimum rain attenuation is found in the months of December and January, which is less than 10 dB compared to all other months.

Since the rain attenuation depends on the frequency, polarization, transmitted frquency, appendix gives three tables of estimated rain attenuation for each polarization in different frequencies up to 40 GHz. The attenuation was predicted to ensure 99.99%, 99.95% and 99.90% availability of the terrestrial microwave link.

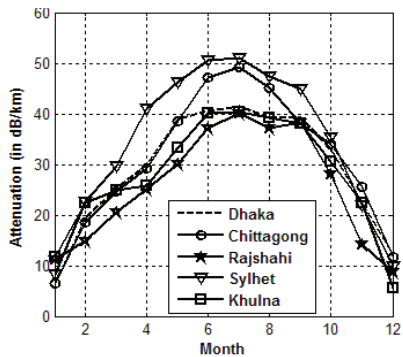

Fig. 10. Variation of attenuation in different months of the year at 10GHz.

## V. CONCLUSIONS

This paper presents the rain attenuation for terrestrial microwave link for the major cities in Bangladesh. The rain rate is maximum at Sylhet among the concerned cities, which is followed by Chittagong, Dhaka, Khulna, and Rajshahi, respectively. The difference between the ITU predicted rain rate and the measured value is about 20%. The highest rainfall amount in a particular month is found at Sylhet, which is almost double of the other cities except Chittagong. The specific attenuation of the mentioned cities remains very close up to 25 GHz, followed by significant difference for higher frequencies. The rain attenuation for longer distance is found comparatively less affected than shorter link length because of non- uniform rain distribution across the link.  It is also found that the horizontally polarized signal is more affected by the precipitation than the circularly and vertically polarized signal. Thus, using vertical polarization in highly rainy area like Sylhet is more economical. The fade margin of horizontally polarized signal required for Rajshahi is smaller than the fade margin for vertically polarized signal at Sylhet. In addition, the amount of fade margin required at Rajshahi to ensure the availability of 99.99 % is almost equal to that for Sylhet to ensure 99.95% availability. Thus, the more availability of terrestrial microwave link can be ensured in Rajshahi, Khulna than the higher rainy areas like Sylhet and Chittagong. The predicted attenuation given in appendix can facilitate engineers in designing terrestrial point-to-point link. All the predicted data can be used to apply frequency diversity technique for reducing the effect of rainfall on microwave communication.

# APPENDIX

*TABLE III: Rain Fade Margin for 40 Km Link in Horizontal Polarization*
Assume: A: 99.99% availability, similarly B: 99.95%, and C: 99.90%

| Frequency (GHz) | Dhaka Availability | | | Chittagong Availability | | | Rajshahi Availability | | | Sylhet Availability | | | Khulna Availability | | |
|---|---|---|---|---|---|---|---|---|---|---|---|---|---|---|---|
| | A | B | C | A | B | C | A | B | C | A | B | C | A | B | C |
| 5 | 4.7 | 2.5 | 1.7 | 5.5 | 2.8 | 2 | 4 | 2 | 1.4 | 6.3 | 3.3 | 2.3 | 4.3 | 2.3 | 1.6 |
| 10 | 32 | 17 | 12 | 36 | 19 | 13 | 29 | 15 | 11 | 40 | 21 | 15 | 31 | 16 | 11 |
| 15 | 63 | 33 | 23 | 69 | 37 | 25 | 57 | 30 | 21 | 76 | 41 | 28 | 60 | 32 | 22 |
| 20 | 94 | 50 | 34 | 102 | 54 | 37 | 85 | 45 | 31 | 112 | 59 | 41 | 90 | 47 | 33 |
| 25 | 122 | 65 | 45 | 133 | 70 | 48 | 111 | 59 | 41 | 144 | 76 | 53 | 117 | 62 | 43 |
| 30 | 145 | 78 | 53 | 159 | 88 | 58 | 134 | 71 | 49 | 172 | 91 | 63 | 141 | 74 | 51 |
| 35 | 167 | 88 | 61 | 180 | 95 | 66 | 154 | 81 | 56 | 194 | 103 | 71 | 160 | 85 | 59 |
| 40 | 183 | 97 | 67 | 197 | 104 | 72 | 169 | 89 | 62 | 212 | 112 | 77 | 176 | 93 | 64 |

*TABLE IV: Rain Fade Margin for 40 Km Link in Vertical Polarization*

| Frequency (GHz) | Dhaka Availability | | | Chittagong Availability | | | Rajshahi Availability | | | Sylhet Availability | | | Khulna Availability | | |
|---|---|---|---|---|---|---|---|---|---|---|---|---|---|---|---|
| | A | B | C | A | B | C | A | B | C | A | B | C | A | B | C |
| 5 | 2.4 | 1.3 | 0.8 | 2.7 | 1.4 | 1 | 2 | 1 | 0.7 | 3 | 1.6 | 1 | 2 | 1 | 0.8 |
| 10 | 25 | 13 | 9 | 27 | 14 | 10 | 22 | 12 | 8 | 30 | 16 | 11 | 23 | 12 | 9 |
| 15 | 48 | 26 | 18 | 53 | 28 | 19 | 44 | 23 | 16 | 58 | 30 | 21 | 46 | 24 | 17 |
| 20 | 70 | 37 | 25 | 76 | 40 | 28 | 64 | 34 | 23 | 82 | 44 | 30 | 67 | 35 | 24 |
| 25 | 94 | 50 | 34 | 101 | 54 | 37 | 86 | 45 | 31 | 110 | 58 | 40 | 90 | 48 | 33 |
| 30 | 118 | 62 | 43 | 127 | 67 | 46 | 108 | 57 | 40 | 137 | 73 | 50 | 113 | 60 | 41 |
| 35 | 139 | 74 | 51 | 150 | 79 | 55 | 128 | 68 | 47 | 161 | 85 | 59 | 134 | 71 | 49 |
| 40 | 157 | 83 | 57 | 168 | 89 | 61 | 145 | 77 | 53 | 181 | 95 | 66 | 151 | 80 | 55 |

*TABLE V: Rain Fade Margin for 40 Km Link in Circular Polarization*

| Frequency (GHz) | Dhaka Availability | | | Chittagong Availability | | | Rajshahi Availability | | | Sylhet Availability | | | Khulna Availability | | |
|---|---|---|---|---|---|---|---|---|---|---|---|---|---|---|---|
| | A | B | C | A | B | C | A | B | C | A | B | C | A | B | C |
| 5 | 3.3 | 1.7 | 1.2 | 3.7 | 2 | 1.3 | 2.8 | 1.5 | 1 | 4.3 | 2.3 | 1.6 | 3 | 1.6 | 1 |
| 10 | 28 | 15 | 10 | 32 | 17 | 12 | 25 | 13 | 9 | 35 | 19 | 13 | 27 | 14 | 10 |
| 15 | 55 | 29 | 20 | 60 | 32 | 22 | 49 | 26 | 18 | 66 | 35 | 24 | 52 | 28 | 19 |
| 20 | 81 | 43 | 29 | 88 | 46 | 32 | 73 | 39 | 27 | 96 | 51 | 35 | 77 | 41 | 28 |
| 25 | 107 | 57 | 39 | 116 | 61 | 42 | 98 | 52 | 36 | 126 | 67 | 46 | 103 | 54 | 37 |
| 30 | 132 | 70 | 48 | 142 | 75 | 52 | 121 | 64 | 44 | 154 | 81 | 56 | 127 | 67 | 46 |
| 35 | 153 | 81 | 56 | 164 | 87 | 60 | 141 | 74 | 51 | 177 | 94 | 65 | 147 | 78 | 54 |
| 40 | 170 | 90 | 62 | 182 | 96 | 66 | 157 | 83 | 57 | 196 | 104 | 71 | 163 | 86 | 60 |